\documentclass[prd,twocolumn,groupedaddress,showpacs,nofootinbib]{revtex4}
\usepackage{graphicx}
\usepackage{dcolumn}
\usepackage{amssymb}
\usepackage{mathrsfs}
\usepackage{amsmath}
\usepackage{epsfig}
\usepackage[dvips]{color}
\usepackage{hhline}

\begin{document}

\title{Baryogenesis and neutron-antineutron oscillation at TeV }

\author{Pei-Hong Gu$^{1}_{}$}
\email{peihong.gu@mpi-hd.mpg.de}

\author{Utpal Sarkar${}^{2,3}$}
\email{utpal@prl.res.in}

\affiliation{ ${}^1$ Max-Planck-Institut f\"{u}r Kernphysik,
Saupfercheckweg 1,
69117 Heidelberg, Germany\\
${}^{2}$ Physical Research Laboratory, Ahmedabad 380009, India\\
${}^3$ McDonnell Center for the Space Sciences, Washington
University, St. Louis, MO 63130, USA}

\begin{abstract}

We propose a TeV extension of the standard model to generate the
cosmological baryon asymmetry with an observable neutron-antineutron
oscillation. The new fields include a singlet fermion, an isotriplet
and two isosinglet diquark scalars. There will be no proton decay
although the Majorana mass of the singlet fermion as well as the
trilinear couplings between one isosinglet diquark and two
isotriplet diquarks softly break the baryon number of two units. The
isosinglet diquarks couple to two right-handed down-type quarks or
to a right-handed up-type quark and a singlet fermion, whereas the
isotriplet diquark couples to two left-handed quarks. The isosinglet
diquarks mediate the three-body decays of the singlet fermion to
realize a TeV baryogenesis without fine tuning the resonant effect.
By the exchange of one singlet fermion and two isosinglet diquarks
and of one isosinglet diquark and two isotriplet diquarks, a
neutron-antineutron oscillation is allowed to verify in the future
experiments.

\end{abstract}


\maketitle

Within the context of the $SU(3)_c^{}\times SU(2)_L^{}\times
U(1)_Y^{}$ standard model (SM), there is a $\textrm{SU(2)}_L^{}$
global anomaly \cite{thooft1976} violating the baryon $(B)$ and
lepton $(L)$ numbers by an equal amount. This anomalous process
becomes fast in the presence of an instanton-like solution, the
sphalerons \cite{krs1985}, during the period $ 100\,\textrm{GeV}
\lesssim T \lesssim 10^{12}_{}\,\textrm{GeV}$. The $B+L$ violating
but $B-L$ conserving sphaleron processes will not affect any
primordial $B-L$ asymmetry and will partially convert the $B-L$
asymmetry to a baryon asymmetry and a lepton asymmetry. So, for a
baryogenesis theory above the weak scale, it should firstly generate
a $B-L$ asymmetry which is composed of a pure baryon asymmetry
\cite{fy2002,bmn2006,gu2007,gs2007} or a pure lepton asymmetry
\cite{fy1986} or any unequal baryon and lepton asymmetries
\cite{gs2007}. For example, the $B-L$ asymmetry in the leptogenesis
\cite{fy1986} scenario is a lepton asymmetry.

In this paper, we propose a new baryogenesis model to generate the
baryon asymmetry at the TeV scale. We extend the SM by four
TeV-scale fields (a singlet fermion, an isotriplet and two
isosinglet diquark scalars). In our model, the baryon number is
softly broken by two units due to the Majorana mass of the singlet
fermion as well as the trilinear couplings between one isosinglet
diquark and two isotriplet diquarks. The isotriplet diquark has the
Yukawa couplings with two left-handed quarks while the isosinglet
diquarks have the Yukawa couplings with two right-handed down-type
quarks. As for the singlet fermion, it has the Yukawa couplings with
one right-handed up-type quark and one isosinglet diquark. The
baryogenesis thus can be realized through the three-body decays of
the singlet fermion which is lighter than the isosinglet diquarks.
In this scenario, we need not fine tune the resonant effect
\cite{fps1995}, like some leptogenesis models \cite{hambye2001}.
Since we have not observed proton decay so far, there is now renewed
interest to look for neutron-antineutron oscillation with the advent
of ultracold neutrons and storage systems. In the presence of the
baryon number violation of two units, our model can avoid the
dangerous proton decay, but result in a neutron-antineutron
oscillation through the exchange of one singlet fermion and two
isosinglet diquarks and of one isosinglet diquark and two isotriplet
diquarks. For the parameter choice of the baryogenesis, the induced
neutron-antineutron oscillation can be sensitive to the forthcoming
experiments.

For simplicity, we do not show the full Lagrangian. Instead, we only
write down the terms relevant to our illustration,
\begin{eqnarray}
\label{lagrangian} \mathcal{L}&\supset& - y_{ai}^{} \delta_a^{}
\bar{u}_{Ri}^{}X_R^c - f_{aij}^{} \delta_a^{} \bar{d}_{Ri}^c
d_{Rj}^{} - h_{ij}^{} \bar{q}_{Li}^c i
\tau_2^{} \Omega q_{Lj}^{} \nonumber\\
&& - \mu_a^{} \delta_a^{}
\textrm{Tr}(\Omega\Omega)-\frac{1}{2}M_X^{}\bar{X}_R^c X_R^{}+
\textrm{H.c.}-M_{\delta_a^{}}^2 \delta_a^\ast\delta_a^{}\nonumber\\
&&-M_\Omega^2 \textrm{Tr}(\Omega^\dagger_{}\Omega) \,.
\end{eqnarray}
Here
\begin{eqnarray}
X_R^{}(\textbf{1},\textbf{1},0)
\end{eqnarray}
is the singlet fermion with a baryon number $B=-1$,
\begin{eqnarray}
\delta(\textbf{3},\textbf{1},\frac{2}{3})\,,\quad
\Omega(\textbf{3},\textbf{3},-\frac{1}{3})=\left[
\begin{array}{rr}
\frac{1}{\sqrt{2}}\omega_{\frac{1}{3}}^{}\quad\quad&\omega_{\frac{2}{3}}^{}\\
[3mm]
\omega_{\frac{4}{3}}^{}\quad\quad&-\frac{1}{\sqrt{2}}\omega_{\frac{1}{3}}^{}
\end{array}\right]
\end{eqnarray}
stand for the isosinglet and isotriplet diquarks with a baryon
number $B=-\frac{2}{3}$, while
\begin{eqnarray}
q_L^{}(\textbf{3},\textbf{2},\frac{1}{6})=\left[
\begin{array}{c}
u_L^{}\\
[3mm] d_L^{}
\end{array}\right]\,,\quad
u_R^{}(\textbf{3},\textbf{1},\frac{2}{3})\,,\quad
d_R^{}(\textbf{3},\textbf{1},-\frac{1}{3})
\end{eqnarray}
denote the SM quarks with a baryon number $B=\frac{1}{3}$. The
baryon number is thus softly broken by the Majorana mass of the
singlet fermion and by the trilinear couplings between the
isosinglet and isotriplet diquarks. Note that the Yukawa couplings
$f$ and $h$ are symmetric for the quark indices, i.e.
$f_{aij}^{}=f_{aji}^{}$ and $h_{ij}^{}=h_{ji}^{}$.
We shall work in the baseis where the Majorana mass
$M_X^{}$ is real so that the Majorana fermion
\begin{eqnarray}
X=X_R^{}+X_R^c=X^c_{}
\end{eqnarray}
can be well defined.

\begin{figure*}
\vspace{11cm} \epsfig{file=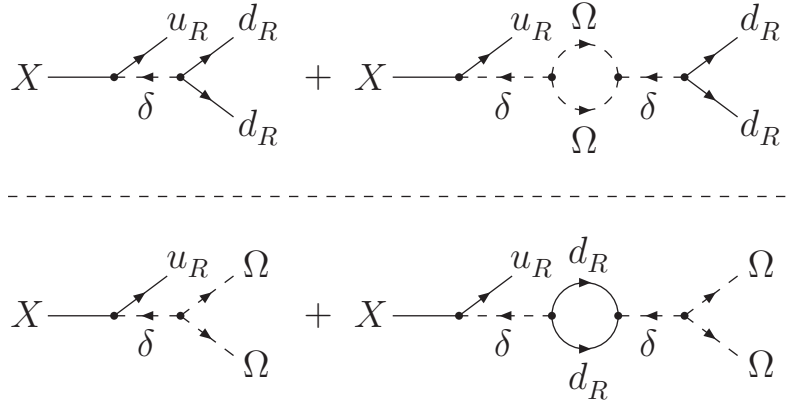, bbllx=4.5cm, bblly=6.0cm,
bburx=14.5cm, bbury=16cm, width=10.cm, height=10.cm, angle=0,
clip=0} \vspace{-15.5cm} \caption{\label{decay} The three-body
decays of the singlet fermion at tree level and one-loop order. The
CP conjugation is not shown for simplicity.}
\end{figure*}

As the singlet fermion and the isosinglet and isotriplet diquarks
are assumed to have the following mass spectrum,
\begin{eqnarray}
2 M_\Omega^{} < M_X^{} < M_\delta^{}\,,
\end{eqnarray}
the decay of the singlet fermion can only be realized by four
three-body modes, i.e.
\begin{eqnarray}
\begin{array}{ll}
X\rightarrow u_R^{}d_R^{}d_R^{}\,,& X\rightarrow
u_R^{}\Omega\Omega\,,
\\
[3mm] X\rightarrow u_R^c d_R^c d_R^c\,, & X\rightarrow
u_R^c\Omega^\ast_{}\Omega^\ast_{}\,, \end{array}
\end{eqnarray}
where the isosinglet diquarks $\delta$ are off-shell. We indicate
the three-body decays at tree level and one-loop order in Fig.
\ref{decay}. For our assignment of the baryon numbers, the decays
$X\rightarrow u_R^{}d_R^{}d_R^{}$ and $X\rightarrow
u_R^c\Omega^\ast_{}\Omega^\ast_{}$ break the baryon number by
$\Delta B=+1$, while the decays $X\rightarrow u_R^c d_R^c d_R^c$ and
$X\rightarrow u_R^c\Omega\Omega$ break the baryon number by $\Delta
B=-1$. So, a baryon asymmetry can be expected if the CP is not
conserved to induce a difference between the decay widths of the
$\Delta B=\pm 1$ processes. We calculate the CP asymmetry at
one-loop order \footnote{In the case with two or more singlet
fermions, we can consider the two-body decays to generate a CP
asymmetry. Like the right-handed neutrinos in the seesaw
\cite{minkowski1977} models, the singlet fermions should have a tiny
mass split to resonantly enhance the CP asymmetry if they are at the
TeV scale. Alternatively, the two isosinglet diquarks can realize
the leptogenesis through their two-body decays even if the singlet
fermion is absent, similar with the isotriplet Higgs scalars
\cite{mz1992,ms1998}. Again, it is necessary for the low scale
isosinglet diquarks to have a fine tuning quasi-degenerate mass
spectrum.},
\begin{eqnarray}
\label{cpasymmetry}
\begin{array}{lcl}\varepsilon_X^{}&=&
\frac{\Gamma_{X\rightarrow
u_R^{}d_R^{}d_R^{}}^{}+\Gamma_{X\rightarrow
u_R^c\Omega^\ast_{}\Omega^\ast_{}}^{}-\Gamma_{X\rightarrow u_R^c
d_R^c d_R^c}^{}-\Gamma_{X\rightarrow
u_R^{}\Omega\Omega}^{}}{\Gamma_{X\rightarrow
u_R^{}d_R^{}d_R^{}}^{}+\Gamma_{X\rightarrow u_R^c d_R^c
d_R^c}^{}+\Gamma_{X\rightarrow
u_R^{}\Omega\Omega}^{}+\Gamma_{X\rightarrow
u_R^c\Omega^\ast_{}\Omega^\ast_{}}^{}}\\
[5mm]
~&=&\frac{3}{2\pi}\frac{\textrm{Im}\left(\sum_{abcijk}^{}y_{ak}^\ast
y_{ck}^{}f_{aij}^{}f_{bij}^\ast\frac{\mu_b^{}\mu_c^\ast
M_X^4}{M_{\delta_a^{}}^2 M_{\delta_b^{}}^2
M_{\delta_c^{}}^2}\right)}{\sum_{abk}^{}y_{ak}^{}y_{bk}^\ast
\left(\sum_{ij}^{}f_{aij}^\ast f_{bij}^{} + 12
\frac{\mu_a^\ast\mu_b^{}}{M_X^2}\right)\frac{M_X^4}{M_{\delta_a^{}}^2
M_{\delta_b^{}}^2}}\,.
\end{array}&&\nonumber\\
&&
\end{eqnarray}
Actually, one can find
\begin{eqnarray}
\begin{array}{cl} ~&\Gamma_{X\rightarrow
u_R^{}d_R^{}d_R^{}}^{}+\Gamma_{X\rightarrow
u_R^c\Omega^\ast_{}\Omega^\ast_{}}^{} \\
[3mm] =&\Gamma_{X\rightarrow u_R^c d_R^c
d_R^c}^{}+\Gamma_{X\rightarrow u_R^{}\Omega\Omega}^{}\,,
\end{array}
\end{eqnarray}
which is guaranteed by CPT conservation and unitary. We also give
the total decay width,
\begin{eqnarray}
\label{decaywidth} \Gamma_X^{}&=&\Gamma_{X\rightarrow
u_R^{}d_R^{}d_R^{}}^{}+\Gamma_{X\rightarrow u_R^c d_R^c
d_R^c}^{}+\Gamma_{X\rightarrow
u_R^{}\Omega\Omega}^{}\nonumber\\
&&+\Gamma_{X\rightarrow
u_R^c\Omega^\ast_{}\Omega^\ast_{}}^{}\nonumber\\
&=&\frac{1}{2^9_{}\pi^3_{}}\sum_{abk}^{}y_{ak}^{}y_{bk}^\ast
\left(\sum_{ij}^{}f_{aij}^\ast f_{bij}^{} + 12
\frac{\mu_a^\ast\mu_b^{}}{M_X^2}\right)\nonumber\\
&&\times\frac{M_X^5}{M_{\delta_a^{}}^2 M_{\delta_b^{}}^2}\,.
\end{eqnarray}
For the following demonstration, we would like to introduce the
parametrization as below,
\begin{eqnarray}
\begin{array}{ll}
~y_{ak}^{}=\bar{y}_{ak}^{}e^{i\alpha_{ak}^{}}_{}\,,&
~\,\alpha_{k}^{}=\alpha_{1k}^{}-\alpha_{2k}^{}\,,\\
[5mm] \,f_{aij}^{}=\bar{f}_{aij}^{}e^{i\beta_{aij}^{}}_{}\,,&
~\beta_{ij}^{}=\beta_{1ij}^{}-\beta_{2ij}^{}\,,\\
[5mm]
\frac{\mu_a^{}}{M_{\delta_a^{}}^{}}=\kappa_a^{}=\bar{\kappa}_a^{}e^{i\gamma_a^{}}_{}\,,&
~~~\gamma=\gamma_{1}^{}-\gamma_{2}^{}\,,\\
[5mm] \frac{M_X^{}}{M_{\delta_a^{}}^{}}=r_a\,,&~
\end{array}
\end{eqnarray}
to specify the CP asymmetry and the decay width by
\begin{eqnarray}
\label{cpasymmetry} \varepsilon_X^{}
=\frac{3}{2\pi}\frac{B}{A}\,,\quad
\Gamma_X^{}=\frac{1}{2^9_{}\pi^3_{}}A M_X^{}
\end{eqnarray}
with
\begin{eqnarray}
A&=&\sum_{k}^{}\bar{y}_{1k}^2(\sum_{ij}^{}\bar{f}_{1ij}^2+12 r_1^2
\bar{\kappa}_1^2)r_1^4\nonumber\\
&&+\sum_{k}^{}\bar{y}_{2k}^2(\sum_{ij}^{}\bar{f}_{2ij}^2+12 r_2^2
\bar{\kappa}_2^2)r_2^4\nonumber\\
&&+2\sum_{k}^{}\bar{y}_{1k}^{}\bar{y}_{2k}^{}[\sum_{ij}^{}\bar{f}_{1ij}^{}
\bar{f}_{2ij}^{}\cos(\alpha_k^{}-\beta_{ij}^{})\nonumber\\
&&+12 r_1^{} r_2^{}
\bar{\kappa}_1^{}\bar{\kappa}_2^{}\cos(\alpha_k^{}-\gamma)]r_1^2
r_2^2\,,\\
B&=& \sum_{ijk}^{}[(r_1^2 \bar{y}_{1k}^2-r_2^2 \bar{y}_{2k}^2)
\bar{f}_{1ij}^{} \bar{f}_{2ij}^{}
\sin(\beta_{ij}^{}-\gamma)\nonumber\\
&&+(r_2^2 \bar{f}_{2ij}^2 - r_1^2
\bar{f}_{1ij}^2)\bar{y}_{1k}^{}\bar{y}_{2k}^{}\sin(\alpha_k^{}-\gamma)]\nonumber\\
&&\times r_1^{}r_2^{}\bar{\kappa}_1^{}\bar{\kappa}_2^{}\,.
\end{eqnarray}

When the Majorana fermions $X$ go out of equilibrium, their CP
violating decays can generate a baryon asymmetry. For example, we
consider the weak washout region, where the out-of-equilibrium
condition can be described by the following quantity,
\begin{eqnarray}
\label{weakwashout}
K=\frac{\Gamma_X^{}}{H}\left|_{T=M_X^{}}^{}\right.\lesssim 1\,.
\end{eqnarray}
Here the Hubble constant $H$ is given by
\begin{eqnarray}
H=\left(\frac{8\pi^{3}_{}g_{\ast}^{}}{90}\right)^{\frac{1}{2}}_{}
\frac{T^{2}_{}}{M_{\textrm{Pl}}^{}}\,,
\end{eqnarray}
with $M_{\textrm{Pl}}^{}=\mathcal{O} (10^{19}_{}\,\textrm{GeV})$
being the Planck mass and $g_\ast^{}=\mathcal{O}(100)$ being the
relativistic degrees of freedom. The induced baryon asymmetry can
approximate to \cite{kt1990}
\begin{eqnarray}
\label{basymmetry} \frac{n_B^{}}{s}\sim
\frac{\varepsilon_X^{}}{g_\ast^{}}\quad \textrm{for}\quad K\lesssim
1\,.
\end{eqnarray}
If the baryogenesis scenario works before the electroweak phase
transition, after which the $B-L$ conserving and $B+L$ violating
sphaleron processes will be highly suppressed, we should require
that other $B-L$ violating interactions (such as the lepton number
violation in the seesaw models \footnote{We should forbid or
suppress the mixing of the singlet fermion to the right-handed
neutrinos in the seesaw model. Otherwise, with the effective Yukawa
couplings of the singlet fermion to the SM lepton and Higgs
doublets, there will be a proton decay. For this purpose, we can
introduce certain discrete symmetries. For example, we can impose a
$Z_2^{}$ symmetry under which the right-handed neutrinos and the SM
leptons are odd while the singlet fermion and diquarks as well as
the SM quarks and Higgs are even. Furthermore, the singlet fermion
can carry a lepton number $L=1$, rather than the present assignment
$B=-1$. In this case, the baryon number violation is changed to be a
$B-L$ violation. The baryon number violation in our model and the
lepton number violation in the seesaw models can be both induced by
a $B-L$ symmetry breaking. If there is a $U(1)_{B-L}^{}$ gauge
symmetry, the singlet fermion should be one of the three
right-handed neutrinos. }) have already decoupled. In the presence
of the sphalerons, the induced baryon asymmetry (\ref{basymmetry}),
which is equivalent to a $B-L$ asymmetry now, will be partially
converted to the final baryon asymmetry,
\begin{eqnarray}
\label{shpaleron} \eta_B^{}=\frac{28}{79}\frac{n_B^{}}{s}\,.
\end{eqnarray}
In the other case that the baryogenesis works after the sphaleron
epoch, the final baryon asymmetry should be just the induced baryon
asymmetry, i.e.
\begin{eqnarray}
\eta_B^{}=\frac{n_B^{}}{s}\,.
\end{eqnarray}

\begin{figure*}
\vspace{12cm} \epsfig{file=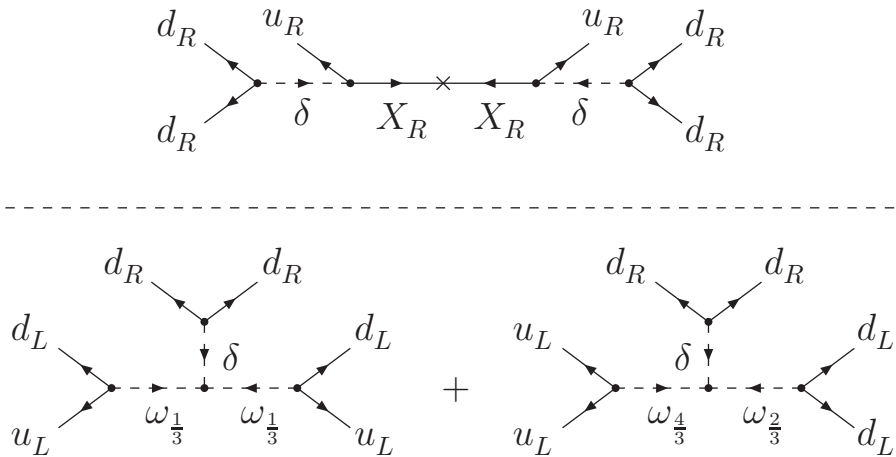, bbllx=4.5cm, bblly=6.0cm,
bburx=14.5cm, bbury=16cm, width=10.cm, height=10.cm, angle=0,
clip=0} \vspace{-14.5cm} \caption{\label{bviolation} The baryon
number violation at low energy. The CP conjugation is not shown for
simplicity.}
\end{figure*}

At low energy, the singlet fermion and the diquarks can mediate the
baryon number violating interactions as shown in Fig.
\ref{bviolation}. The effective operators should be
\begin{eqnarray}
\mathcal{L}_{eff}^{\Delta B =2}&=&-\sum_{ab}^{}\frac{ f_{aij}^\ast
y_{ak}^{} y_{bl}^{}f_{bmn}^\ast}{M_X^{}M_{\delta_a^{}}^2
M_{\delta_b^{}}^2}\bar{d}_{Ri}^{}d_{Rj}^c
\bar{u}_{Rk}^{}u_{Rl}^c\bar{d}_{Rn}^{}d_{Rm}^c \nonumber\\
&&-\sum_{a}^{}\frac{4\mu_a^{}f_{aij}^\ast h_{kl}^\ast
h_{mn}^\ast}{M_{\delta_a^{}}^2 M_\Omega^4}\bar{d}_{Ri}^{}d_{Rj}^c
(\bar{u}_{Lk}^{}d_{Ll}^c\bar{u}_{Lm}^{}d_{Ln}^c\nonumber\\
&& - \frac{1}{2}
\bar{u}_{Lk}^{}u_{Ll}^c\bar{d}_{Lm}^{}d_{Ln}^c)+\textrm{H.c.}\,,
\end{eqnarray}
where the first term is mediated by one singlet fermion and two
isosinglet diquarks while the second term is mediated by one
isosinglet diquark and two isotriplet diquarks. From the above
$\Delta B =\pm 2$ interactions, we can easily read the operators for
the neutron-antineutron oscillation,
\begin{eqnarray}
\label{nbarn}
\mathcal{L}_{eff}^{n-\bar{n}}&=&-\sum_{ab}^{}\frac{f_{a11}^\ast
y_{a1}^{} y_{b1}^{}f_{b11}^\ast}{M_X^{}M_{\delta_a^{}}^2
M_{\delta_b^{}}^2} \bar{d}_{R}^{}d_{R}^c
\bar{u}_{R}^{}u_{R}^c\bar{d}_{R}^{}d_{R}^c\nonumber\\
&&-\sum_{a}^{}\frac{6\mu_a^{}f_{a11}^\ast h_{11}^\ast
h_{11}^\ast}{M_{\delta_a^{}}^2 M_\Omega^4}\bar{d}_{R}^{}d_{R}^c
\bar{u}_{L}^{}d_{L}^c\bar{u}_{L}^{}d_{L}^c
\nonumber\\
&&+\textrm{H.c.}\,.
\end{eqnarray}

We now indicate that our model can simultaneously generate a desired
baryon asymmetry and an accessible neutron-antineutron oscillation.
For simplicity, we take
\begin{eqnarray}
&&\bar{y}_{1k}^{}=\bar{y}_{2k}^{}=\tilde{y}\,,~~\bar{f}_{1ij}^{}=\bar{f}_{2ij}^{}=\tilde{f}\,,~~
\bar{\kappa}_1^{}=\bar{\kappa}_2^{}=\tilde{\kappa}\,,\nonumber\\
&&\gamma-\alpha_{k}^{}=\beta_{ij}^{}-\gamma=\delta
\end{eqnarray}
to derive
\begin{eqnarray}
\begin{array}{l}
\varepsilon_X^{}=
\frac{9}{\pi}\frac{\tilde{f}^2_{}\tilde{\kappa}^2_{}r_1^{}r_2^{}(r_1^2-r_2^2)\sin\delta}
{3\tilde{f}^2_{}(r_1^4+2r_1^2r_2^2\cos2\delta+r_2^4)+4\tilde{\kappa}^2_{}(r_1^6+2r_1^3r_2^3\cos\delta+r_2^6)}
\,,
\end{array}
\end{eqnarray}
as well as
\begin{eqnarray}
K&=&
\frac{3^3_{}\sqrt{5}}{2^{10}_{}\pi^{\frac{9}{2}}_{}\sqrt{g_\ast^{}}}\frac{M_{\textrm{Pl}}^{}}{M_X^{}}
\tilde{y}^2_{}[3\tilde{f}^2_{}(r_1^4+2r_1^2r_2^2\cos2\delta + r_2^4)\nonumber\\
&&+4\tilde{\kappa}^2_{}(r_1^6+2r_1^3r_2^3\cos\delta+r_2^6)]\,.
\end{eqnarray}
The singlet fermion and the diquarks are taken at the TeV scale such
as
\begin{eqnarray}
\begin{array}{ll}
\,M_\Omega^{} = 0.3\,\textrm{TeV}\,,&M_X^{}=
1\,\textrm{TeV}\,, \\
[5mm] M_{\delta_1^{}}^{}= 3\,\textrm{TeV}\,,&M_{\delta_2^{}}^{}=
3.3\,\textrm{TeV}\,.
\end{array}
\end{eqnarray}
With the leading
\begin{eqnarray}
r_1^{}\simeq 0.33\,,~~r_2^{}\simeq 0.3\,,
\end{eqnarray}
we can obtain
\begin{eqnarray}
\varepsilon_X^{}\simeq 2.9\times 10^{-8}_{}\,, \quad K \simeq
0.18\,,
\end{eqnarray}
by further inputting
\begin{eqnarray}
\tilde{y}=\tilde{f}=\tilde{\kappa} = 1.5\times 10^{-3}_{}\,,~~
\sin\delta= 0.5\,.
\end{eqnarray}
The final baryon asymmetry determined by Eq. (\ref{shpaleron}) can
explain the measured value,
\begin{eqnarray}
\eta_B^{}\sim 10^{-10}_{}\,.
\end{eqnarray}
At the same time, the neutron-antineutron oscillation described by
the first term of Eq. (\ref{nbarn}) can be observed in the future
since its strength is of the order of
\begin{eqnarray}
G^{u_R^{}d_R^{}d_R^{}}_{n-\bar{n}}\sim
10^{-28}_{}\textrm{GeV}^{-5}_{}\,,
\end{eqnarray}
which is close to the currently experimental bound
\cite{takita1986}. As for the neutrino-antineutrino oscillation from
the second term of Eq. (\ref{nbarn}), its strength can also arrive
at the same magnitude, i.e.
\begin{eqnarray}
G^{u_L^{}d_L^{}d_R^{}}_{n-\bar{n}}\sim
10^{-28}_{}\textrm{GeV}^{-5}_{}\quad\textrm{for}\quad h_{11}^{}\sim
10^{-5}\,.
\end{eqnarray}

In this paper, we extended the SM by a singlet fermion, an
isotriplet diquarks and two isosinglet diquarks to generate the
cosmological baryon asymmetry with a testable neutron-antineutron
oscillation. The new fields are all at the TeV scale. So, they can
be verified at colliders (such as the LHC) because the diquarks can
be produced through their gauge interactions and then can decay into
the quarks and the singlet fermion.

\textbf{Acknowledgement}: PHG is supported by the Alexander von
Humboldt Foundation. US thanks R. Cowsik for arranging his visit
as the Clark Way Harrison visiting professor.

\end{document}